# A Convergent control strategy for quantum systems

Shuang Cong*, Yuesheng Lou, Jianxiu Liu, Sen Kuang

Department of Automation, University of Science and Technology of China,

Hefei, 230027, China
*Corresponding author. Phone +86-551-63600710. Fax +86-551-63603244

**Abstract:** In the interaction picture, a sufficient and necessary condition that guarantees the convergence of closed quantum control system is proposed in this paper. Theoretical derivation and the proof show that it is possible to achieve the convergence to the target state by constructing an observable operator in an energy function and selecting control Hamiltonians. Numerical simulation experiments on a four-level system verify the effectiveness of the proposed control strategy.

## 1. Introduction

The Lyapunov-based control method has played a significant role in classical system control. Originally used in feedback control to analyze the stability or convergence of the controlled system, the Lyapunov-based control methods have be successfully used in the quantum control systems [1]-[7]. The controller designed by means of the Lyapunov method usually is only a stable controller. A convergent controller needs more conditions. The research results obtained have indicated that some requirements on the Hamiltonian and the target stateare needed [6].

The most important contribution of this paper is to study the convergence to an arbitrary target state of quantum control system based on the Lyapunov method. By transforming the system to be controlled into the interaction picture, the complexity is reduced but the system transformed becomes a non-autonomous system to which the LaSalle invariance principle can not be used. In order to be able to analyze the convergence of the control system, we introduce the Barbalat lemma in this paper which is most different from the method used in [6]. Another key point of this paper is to introduce the concept of energy function which extends the context of the Lyapunov function. The energy function needs not any more limitation of positive semi-definite, it can be negative value, which cannot affect the stability of system but does the convergence. We also introduce an observable operator $P$ into the energy function [8][9], which provides a freedom of designing convergent controller.

The remainder of the paper is organized as follows. In Sec. 2, the control problem is described and control laws are designed in the interaction picture. In Sec.3, the main results about the construction of observable operator and the convergence conditions of system control are presented. The proof of the convergence conditions is derived in Sec. 4. In Sec. 5, a four-level quantum system is used to verify the proposed method by simulation experiments. The conclusions with brief remarks are made in Sec. 6.

## 2. Problem description

Consider a closed quantum system with dimension $n$:

$$i\hbar\dot{\rho} = [H(t), \rho] \tag{1a}$$

$$H(t) = H_0 + \sum_{j=1}^{} H_j u_j(t) \tag{1b}$$

where $H_0$ is a free Hamiltonian, which is a Hermite diagonal matrix; $H_j$ is a control Hamiltonian; $u_j$ is a control field, and $\rho$ is the state density matrix of the system, which is Hermite, too. By transforming system (1) into the interaction picture, viz. setting $\rho' = e^{iH_0 t/\hbar} \rho e^{-iH_0 t/\hbar}$, the Liouville equation of the closed quantum system after transformation can be written as

$$\dot{\rho}' = [\sum_j A_j(t) u_j(t), \rho'] \tag{2a}$$

$$A_j(t) = e^{iH_0 t/\hbar} H_j e^{-iH_0 t/\hbar} / i\hbar \tag{2b}$$

where $A_j(t)$ is the control Hamiltonian in the interaction picture. It can be verified that such a transformation does not change the population distribution of the system state. Therefore $|\rho\rangle$ is equivalent to $|\rho'\rangle$ and we will omit the label "'" in the following text for convenience.

Define a bounded energy function $E(\rho)$ by:

$$E(\rho) = tr(P\rho) \tag{3}$$

where $P$ is an observable operator without the limitation of positive definite.

The control goal is to steer an arbitrary initial state in system (2) to the desired target state which can be eigenstate, superposition state, or mixed state by designing a convergent control law. Clearly, one could achieve such a goal by associating the minimum point of energy function (3) with target state $\rho_f$ and guaranteeing its monotonical decrease. Let its first order time derivative be negative semi-definite:

$$\dot{E}(\rho(t)) = \sum_j u_j tr([\rho(t), P] A_j) \leq 0 \tag{4}$$

Then the following control law can be derived:

$$u_j(t) = -\kappa_j(t) tr([\rho(t), P] A_j(t)), \kappa_j(t) > 0 \tag{5}$$

where $\kappa_j(t)$ is a control gain, which used to adjust the convergent speed of system.

Generally, control law (5) is only a stable control. The system can only converge to a local extreme point of some system trajectories instead of the global minimum point. In order to solve such a problem, an observable operator $P$ will be designed and conditions on control Hamiltonians will be proposed in this paper under certain assumptions.

## 3. Main results

In this section we'll first give an appropriate construction of the observable operator $P$, which makes $\rho_f$ be the minimum point of (3), and then give conditions on control Hamiltonians such that control law becomes convergent.

**Theorem 1** [10] Suppose system (2) is transferred from an given initial state to the desired target state $\rho_f$, whose coherent vector is $R_{\rho_f} = (f_1, f_2, \cdots, f_{n^2-1})^T$. In order to make the target state $\rho_f$ be the minimum point of the energy function (3), the coherent vectors of $P$ and $\rho_f$ must have opposite directions, i.e.,

$$R_P = \lambda R_{\rho_f}, \lambda < 0 \tag{6}$$

in which $\rho_f = I/n + \sum_j f_j X_j$ and $P = c_0 I + \sum_k c_k X_k$, where $\{-i\mathbf{X}_l\}$ is the basis of $su(n)$, which satisfies $tr(\mathbf{X}_l \mathbf{X}_j) = \delta_{lj}$; $x_l$ is a real number. The vector $R_\rho$ formed by $x_1, x_2, \cdots,$ and $x_{n^2-1}$ is called the coherent vector of density matrix $\rho$, i.e., $R_\rho = (x_1, x_2, \cdots, x_{n^2-1})^T$.

Theorem 1 implies that $P$ is not unique. For instance, the simplest way to get observable operator $P$ is to let $P = -\rho_f$. For a two-level system with the target state $\rho_f = |0\rangle\langle 0|$, $R_P$ and $R_{\rho_f}$ have opposite directions when one takes $P = a|0\rangle\langle 0| + b|1\rangle\langle 1|$ with $a < b$. Generally, for pure states, the following conclusion holds:

**Corollary 1** Suppose the system state and the target state are both pure states, i.e., $\rho = |\psi\rangle\langle\psi|$ and $\rho_f = |\psi_f\rangle\langle\psi_f|$. In order for $\rho_f$ to be the minimum point of (3), observable operator $P$ can be constructed by an orthonormal basis containing target state $\rho_f$:

$$P = p_h \sum_j |\psi_j\rangle\langle\psi_j| + p_l |\psi_f\rangle\langle\psi_f| \quad , \text{ for } \langle\psi_j|\psi_f\rangle = \delta_{jk} \text{ and } p_l < p_h \tag{7}$$

**Remarks**: It can be proven from (7) that energy function (3) reaches its minimum if and only if $|\psi\rangle = |\psi_f\rangle$ for a two-level system with pure states. According to Theorem 1, the coherent vector of $P$ in (7) has an opposite direction with that of $\rho_f$, i.e. $E(\rho) = -tr(\rho\rho_f) \leq 0$. This energy function is not a standard Lyapunov function with $E(\rho) \geq 0$ any more because $P$ is no longer positive definite. That is the reason why we define a bounded energy function (3).

Theorem 1 and Corollary 1 are the construction methods and the expression of observable operator $P$ for general states including the pure states. Such an observable operator guarantees that target state $\rho_f$ is the minimum point of energy function (3) and control law (5) is stable. In order to make such controls be convergent, some restrictions on control Hamiltonians should be placed. To this end, three assumptions are given in advance.

**Assumption 1** System (1) is strongly regular, i.e., all the transition frequencies are different, viz. $\Delta_{jk} \neq \Delta_{pq}$, $(j,k) \neq (p,q)$, where $\Delta_{jk} = \lambda_j - \lambda_k$ and $e_j$ is an eigenvalues of $H_0$.

**Assumption 2** Control Hamiltonian $H_l$ has a particular structure: $H_l \in \{\hbar h_{jk} \mid h_{jk} = |j\rangle\langle k| + |k\rangle\langle j|, j > k\}$, where $|j\rangle$ is the eigenstate associated with $\lambda_j$.

**Assumption 3** Initial state and target state are unitarily equivalent, i.e., there exists a unitary transform $U$ such that $\rho_f = U\rho U^\dagger$.

**Remarks**: Assumption 1 indicates that transition frequencies between all the energy levels are distinct, so that each transition can be distinguished and selected. This enables Assumption 2 to hold by considering rotating wave approximation. Operator $h_{jk}$ denotes that the transition between energy levels $j$ and $k$ is admitted. If two transitions are the same, e.g., $\Delta_{12} = \Delta_{34}$, then $h_{12}$ and $h_{34}$ should be incorporated as $h = h_{12} + h_{34}$.

Now we'll propose a sufficient and necessary condition that guarantees the convergence of control system (1):

**Theorem 2** Based on Assumptions 1, 2 and 3, for any given initial and the target states, observable operator $P$ is constructed according to (6). Then, a sufficient and necessary condition that system (1) with control law (5) converges to the target state is: for $\forall j, k$, $\exists l$, s.t. $H_l = \hbar h_{jk}$.

## 4. Proof of convergence

Theorem 2 will be proven in this section by analyzing the states in the invariant set.

If the controlled system is asymptotically stable, the system will converge to the target state. For autonomous systems, asymptotically stability can be analyzed by the LaSalle principle [11]. But in the present situation, the controlled system becomes non-autonomous in the interaction picture. However, one can get an analogous conclusion by the Barbalat lemma:

**Lemma 1:** If scalar function $E(x,t)$ satisfies: (1) $E(x,t)$ is lower bounded; (2) $\dot{E}(x,t)$ is negative semi-definite; (3) $\dot{E}(x,t)$ is uniformly continuous in time. Then, $\dot{E}(x,t) \to 0$ as $t \to \infty$.

For the pure state, the energy function $E(\rho) = tr(P\rho_f) = -tr(\rho\rho_f) \geq -tr(\rho_f^2)$ is lower bounded. Its first derivative is negative semi-definite under control law (5). The second derivative is

$$\ddot{E}(\rho,t) = \sum_j \{u_j tr([\dot{\rho}(t), P]A_j(t)) + u_j tr([\rho(t), P]\dot{A}_j(t))\} \quad (8)$$

Eq. (8) is bounded when the inputs are bounded. Thus, $\dot{E}(\rho,t)$ is uniformly continuous in time. According to Lemma 1, the first order time derivative of the energy function converges to zero, viz., $\dot{E}(\rho(\infty),\infty) = 0$.

Let $\mathcal{R}$ be the set of critical points on any dynamic trajectory, viz.,

$$\mathcal{R} \equiv \{\rho : tr([\rho,P]A_j(t)) = 0, \forall j, t\} \tag{9}$$

Then, the controlled system converges to $\mathcal{R}$.

Moreover, system (2) is homogeneous, and the states in $\mathcal{R}$ make their controls be zero according to (5). So, the largest invariant set in $\mathcal{R}$ is itself.

**Theorem 3:** System (2) with control law (5) converges to the set $\mathcal{R}$ defined by (9).

Wang and Schirmer [6] indicated that kinematically critical points $\rho_s$ can be defined by $[\rho_s, \rho_f] = 0$, and if the system is controllable, then all the critical points except the maximum and minimum points are kinematically critical stable. However, kinematical analysis implicates that the direction fields of state evolution are unrestricted. In fact, the direction fields determined by the Hamiltonians are usually non-arbitrary. This means that some trajectories in kinematics are forbidden in dynamics. That is to say, the results in kinematics and in dynamics may be different. So, dynamical stability of critical points in $\mathcal{R}$ will be further analyzed as follows.

First, we show that if the conditions in Theorem 2 are satisfied, then critical points in kinematics and in dynamics are the same.

**Lemma 2:** If the condition in Theorem 2 is satisfied. Then, $\rho_s$ can be defined by $[\rho_s, P] = D$, viz., the invariant set $\mathcal{R}$ in Theorem 3 can be redefined by:

$$\mathcal{R} \equiv \{\rho : [\rho, P] = D\} \tag{10}$$

where $D$ is a diagonal matrix.

**Lemma 3:** [12] if $rank\tilde{A}(\vec{P}) = n^2 - n$ holds, then the invariant set $\mathcal{R}$ is regular, viz. $\mathcal{R} \equiv \{\rho_s : [\rho_s, P] = 0\}$, where one denotes $[\rho, P] = Ad_p(\vec{\rho})$, and $Ad_p$ is a linear map from Hermitian or anti-Hermitian matrices into $su(n)$. Let $A(\vec{P})$ be the real $(n^2-1)*(n^2-1)$ matrix corresponding to the Bloch representation of $Ad_p$. Denote $su(n) = T \oplus C$ and $R^{n^2-1} = S_T \oplus S_C$, where $S_C$ and $S_T$ are real subspaces corresponding to the Cartan and non-Cartan subspaces, $C$ and $T$, respectively. And then $\tilde{A}(\vec{P})$ be the first $n^2 - n$ rows of $A(\vec{P})$.

For target state $\rho_f$ with diagonal type, the inviariant set with $P = -\rho_f$ in (10) can be simplified as:

$$\mathcal{R} \equiv \{\rho : [\rho, P] = 0\} \tag{11}$$

For non-diagonal target state $\rho_f$ included some mixed states and superposition states, Lemma 3 is employed. If $\rho_f$ satisfies Lemma 3 automatically, one obtains (11) with $P = -\rho_f$. Otherwise, taking superposition states for example, $P$ is chosen as Eq. (7). We can always choose eigenvalues $p_i$ to meet the condition of Lemma 3, then (11) is obtained. To sum up, for most diagnal target states contained diagonal target states, the superposition states and some mixed-states satisfy Lemma 3, there are two ways to construct $P$: one is $P = -\rho_f$ and the other one is (7), by which the invariant set can be reduced to (11). All the convergence analysis will be carried out based on (11).

Next, we will give a sufficient and necessary condition on control Hamiltonians such that all the states in $\mathcal{R}$ except the maximum and minimum points are dynamically critical stable under Assumptions 1 and 2.

For a dynamically minimum point, $\ddot{E}(\rho_s, t) \geq 0$ holds for any time and any control $u = (u_1, u_2, \cdots, u_N)$. Otherwise, the state will be critical stable. That is to say, it is enough to consider the second derivative of the energy function.

For $tr([\rho_s, P]\dot{A}_j(t)) = \frac{i}{\hbar} tr([\rho_s, P][H_0, A_j]) = 0,$ the following equation holds:

$$\ddot{E}(\rho_s,t) = tr\left(\sum_j u_j \dot{A}_j(t)[\rho_s,P]\right) + tr\left(\left[\sum_j u_j A_j(t),\rho_s\right]\left[P,\sum_j u_j A_j(t)\right]\right)$$

$$= tr\left(\left[\sum_j u_j A_j(t),\rho_s\right]\left[P,\sum_j u_j A_j(t)\right]\right)$$

If $P = -\rho_f$, one has:

$$\ddot{E}(\rho_s,t) = tr([\sum_j u_j A_j(t),\rho_s][\sum_j u_j A_j(t),\rho_f]) \tag{12}$$

If $P$ is chosen as (7), where we denote $\rho_h = |\psi_h\rangle\langle\psi_h|$, one obtains:

$$\ddot{E}(\rho_s,t) = p_l tr\left(\left[\sum_j u_j A_j(t),\rho_s\right]\left[\sum_j u_j A_j(t),\rho_f\right]\right) + \sum_h p_h tr\left(\left[\sum_j u_j A_j(t),\rho_s\right]\left[\sum_j u_j A_j(t),\rho_h\right]\right)$$

$$= p_l tr\left(\left[\sum_j u_j A_j(t),\rho_s\right]\left[\sum_j u_j A_j(t),\rho_f\right]\right)$$

(13)

Obviously, $\rho_f$ is a stable point since $\ddot{E}(\rho_f,t) \geq 0$ holds for $\forall t$. Further analysis implicates the following Lemma:

**Lemma 4:** For any diagonal target state $\rho_f$ a sufficient and necessary condition such that all the states in $\mathcal{R}$ except the minimum point are dynamically critical stable is: for $\forall l_0, r_0$, $\exists A_q(t)$, such that $(A_q(t))_{l_0 r_0} \neq 0$.

**Proof:** Necessity. It is equivalent to prove that if $\exists l_0, r_0$, such that $(A_q(t))_{l_0 r_0} = 0$ for $A_q(t), (A_q(t))_{lr} \neq 0$, then $\exists \rho_f, \rho_s \neq \rho_f$ such that $\ddot{E}(\rho_s,t) \geq 0$ holds for $\forall t$ and any controls.

We can denote diagonal matrices $\rho_f$ and $\rho_s$ as follows. Let the two smallest elements of $\rho_f$ are $(\rho_f)_{r_0 r_0}$ and $(\rho_f)_{l_0 l_0}$. Then, $\rho_s$ can be selected by exchanging the $l_0$ th and $r_0$ th elements of $\rho_f$, i.e., $(\rho_s)_{l_0 l_0} = (\rho_f)_{r_0 r_0}$, and $(\rho_s)_{r_0 r_0} = (\rho_f)_{l_0 l_0}$. Obviously, $\rho_f$ and $\rho_s$ are commutable, viz., $\rho_s \in \mathcal{R}$.

According to Assumption 1 and 2, one has:

$$i\hbar A_q(t) \in \left\{ |j\rangle\langle k| e^{i\omega_{jk}t/\hbar} + |k\rangle\langle j| e^{-i\omega_{jk}t/\hbar}, j < k \right\} \quad (14)$$

Denote $H_q$ by $H_{lr}$ and $u_q$ by $u_{lr}$. If $(A_q(t))_{lr} \neq 0$, it can be obtained from (12) that

$$\ddot{E}(\rho_s) = \sum_{l \neq l_0, r \neq r_0} u_{lr}^2 \left( (\rho_s)_{rr} - (\rho_s)_{ll} \right) \left( (\rho_f)_{rr} - (\rho_f)_{ll} \right) \quad (15)$$

For $\forall j \neq l_0, j \neq r_0$, $(\rho_s)_{jj} = (\rho_f)_{jj}$, $\left( (\rho_s)_{rr} - (\rho_s)_{ll} \right) \left( (\rho_f)_{rr} - (\rho_f)_{ll} \right) \geq 0$ holds. Therefore, $\ddot{E}(\rho_s, t) \geq 0$ holds for $\forall t$ and any controls. The necessity is proven.

Sufficiency. It is enough to prove that $\exists u$, $\ddot{E}(\rho_s) < 0$.

For $[\rho_s, \rho_f] = 0$, $\rho_s$ is a diagonal matrix, too. Since $\rho_f$ and $\rho_s$ are unitarily equivalent, $\rho_s$ has the same spectrum with $\rho_f$, that is, the diagonal elements of $\rho_s$ is a permutation of that of $\rho_f$. Suppose the largest element of the permutated ones is $(\rho_f)_{l_0 l_0}$ or $(\rho_s)_{r_0 r_0}$, then we have $(\rho_s)_{r_0 r_0} - (\rho_s)_{l_0 l_0} > 0$ and $(\rho_f)_{r_0 r_0} - (\rho_f)_{l_0 l_0} < 0$. According to (15), if $u_{l_0 r_0} \neq 0$ and other controls are equal to zero, then $\ddot{E}(\rho_s) < 0$. The sufficiency is proven.
Lemma 4 is proven.

For a general $\rho_f$, the proof of the necessity is the same as Lemma 4, one needs only to prove the sufficiency.

Now, suppose $\rho_f = \sum_j c_j |\psi_j\rangle\langle\psi_j|$, where $|\psi_j\rangle = \sum_k \alpha_{jk} |k\rangle$. There must be $U = \sum_j |j\rangle\langle\psi_j|$ to diagonalize $\rho_f$ and $\rho_s$ simultaneously, then

$$\begin{aligned}\ddot{E}(\rho_s) &= tr([\sum_j u_j U A_j(t) U^\dagger, U\rho_s U^\dagger][\sum_j u_j U A_j(t) U^\dagger, U\rho_f U^\dagger]) \\ &\equiv tr([\sum_j u_j U A_j(t) U^\dagger, D_s][\sum_j u_j U A_j(t) U^\dagger, D_f])\end{aligned} \quad (16)$$

where $D_s, D_f$ are two diagonal matrices. Especially, if $\rho_f$ is a superposition state, $P$ is chosen as (7). Eq. (16) is also obtained from (13). Calculate $\sum_j u_j UA_j(t)U^\dagger$ one gets

$$\sum_j u_j UA_j(t)U^\dagger = -i\sum_j 2\operatorname{Re}\{(\sum_{l<r} u_{lr}\alpha_{jl}^*\alpha_{jr}e^{i\omega_{lr}t})|j\rangle\langle j|\}$$
$$-i\sum_{j<k}\{[\sum_{l<r}u_{lr}(\alpha_{jl}^*\alpha_{kr}e^{i\omega_{lr}t}+\alpha_{jr}^*\alpha_{kl}e^{-i\omega_{lr}t})]|j\rangle\langle k| \quad (17)$$
$$+[\sum_{l<r}u_{lr}(\alpha_{kl}^*\alpha_{jr}e^{i\omega_{lr}t}+\alpha_{kr}^*\alpha_{jl}e^{-i\omega_{lr}t})]|k\rangle\langle j|\}$$

in which, the first sum on the right hand side is a diagonal matrix, which commutes with $D_s, D_f$. So, it can be ignored. Let

$$\begin{aligned} f_{jk} &\equiv \sum_{l<r} u_{lr}(\alpha_{jl}^*\alpha_{kr}e^{i\omega_{lr}t}+\alpha_{jr}^*\alpha_{kl}e^{-i\omega_{lr}t}) \\ &= \langle\psi_j|e^{iH_0t}(\sum_{l<r}u_{lr}H_{lr})e^{-iH_0t}|\psi_k\rangle \\ &= \langle\psi'_j|(\sum_{l<r}u_{lr}H_{lr})|\psi'_k\rangle \end{aligned} \quad (18)$$

Then

$$\ddot{E}(\rho_s) = \sum_{j,k}|f_{jk}|^2((D_s)_{kk}-(D_s)_{jj})((D_f)_{kk}-(D_f)_{jj}) \quad (19)$$

Eq. (18) can be rewritten as:

$$\begin{pmatrix} \langle\psi'_1|H_{12}|\psi'_2\rangle & \langle\psi'_1|H_{13}|\psi'_2\rangle & \cdots & \langle\psi'_1|H_{(n-1)n}|\psi'_2\rangle \\ \langle\psi'_1|H_{12}|\psi'_3\rangle & \langle\psi'_1|H_{13}|\psi'_3\rangle & \cdots & \langle\psi'_1|H_{(n-1)n}|\psi'_3\rangle \\ \vdots & \vdots & \ddots & \vdots \\ \langle\psi'_{n-1}|H_{12}|\psi'_n\rangle & \langle\psi'_{n-1}|H_{13}|\psi'_n\rangle & \cdots & \langle\psi'_{n-1}|H_{(n-1)n}|\psi'_n\rangle \end{pmatrix}\begin{pmatrix} u_{12} \\ u_{13} \\ \vdots \\ u_{(n-1)n} \end{pmatrix} = \begin{pmatrix} f_{12} \\ f_{13} \\ \vdots \\ f_{(n-1)n} \end{pmatrix} \quad (20)$$

If the square matrix in (20) is noted as $M$, it is easy to verify that if the condition in Theorem 2 is satisfied, $M$ has full rank. Because if there exists a linear combination of some columns in the square matrix being zero, then for $\forall j,k$, there is

$$\langle\psi'_j|(\sum_{l<r}\beta_{lr}H_{lr})|\psi'_k\rangle = 0 \quad (21)$$

Consider arbitrary two states $|\psi'_j\rangle$ and $|\psi'_k\rangle$, equation (21) holds only for $\beta_{lr}=0$, viz. the matrix $M$ is full rank.

In the following we will find the control $u$ that make $\ddot{E}(\rho_s,u)<0$. Equation (21) can be rewriten as :

$$\ddot{E}(\rho_s,u) = f^\dagger K f = u^\dagger M^\dagger K M u \qquad (22)$$

in which K is a diagonal matrix and defined by:

$$K = \begin{pmatrix} ((D_s)_{22}-(D_s)_{11})((D_f)_{22}-(D_f)_{11}) & \cdots & 0 \\ \vdots & \ddots & \vdots \\ 0 & \cdots & ((D_s)_{nn}-(D_s)_{(n-1)(n-1)})((D_f)_{nn}-(D_f)_{(n-1)(n-1)}) \end{pmatrix} \qquad (23)$$

Matrix K has at least one negative element, because the two matrices $D_s$ and $D_f$ satisfy $D_f \neq D_s$, and their diagonal elements are ranges of the same numbers. In other words, matrix K is non-positive definite. Therefore the matrix $M^\dagger K M$ is non-positive definite, too. And its eigenvalues includ at least one real negative element. Viz.

$$M^\dagger K M = T^\dagger Q T \qquad (24)$$

where $Q$ is a real negative diagonal matrix in its $k_0$th element; $T$ is an unitary matrix whose $k_0$th row is supposed to be $z^\dagger$. The matrix $T$ and the vector $z$ can be separated into real and imaginary parts, respectively, as:

$$T = A + iB \qquad (25)$$

and

$$z = x + iy \qquad (26)$$

Because $\ddot{E}(\rho_s,u)$ is a real number, so the imaginary part of the matrix $M^+ K M$ has no contribution to $\ddot{E}(\rho_s,u)$, thus equation (27) can be written as:

$$\ddot{E}(\rho_s,u) = u^+ (A'QA + B'QB) u \qquad (27)$$

If the control $u$ is replaced by $z$:

$$\begin{aligned}
\ddot{E}(\rho_s, z) &= (Q)_{k_0 k_0} \\
&= z^{\dagger}(A'QA + B'QB)z \\
&= x'(A'QA + B'QB)x + y'(A'QA + B'QB)y \\
&< 0
\end{aligned} \quad (28)$$

Then one can see that at least one of the two items $x'(A'QA + B'QB)x$ and $y'(A'QA + B'QB)y$ is negative. It is no harm to suppose that $x'(A'QA + B'QB)x < 0$, then if only the control is the real part of the vector $z$, one can has $\ddot{E}(\rho_s, \mathcal{R}e(z)) < 0$. So if the condition in Theorem 2 is satisfied, all the states in the largest invariant set $\mathcal{R}$ except the target state are dynamically critical stable, i.e., Theorem 2 can be obtained.

Theorem 2 means that any restriction on the evolution direction will produce new dynamically stable points, which may have a strong impact on control design. The condition in Theorem 2 is obtained based on Assumption 2, which in fact defines a basic orthonormal basis for permissible state evolving directions. This basic basis spans a full space of evolving direction fields when the condition in theorem 2 is satisfied. In fact, Assumption 2 is not necessary, if another basis spans the same space of evolving direction fields, then all the states in invariant set $\mathcal{R}$ except the target state are dynamically critical stable, and then control law (5) is still convergent.

## 5. Numerical simulation examples and discuss

Numerical examples are simulated on a four-level quantum system in this section to verify the effectiveness of the proposed control strategy. The free Hamiltonian of the controlled system $H_0$ is:

$$H_0 = \sum_{j=1}^{n} E_j |j\rangle\langle j| \quad (29)$$

where $E_1 = 0.4948, E_2 = 1.4529, E_3 = 2.3691$ and $E_4 = 3.2434$. Their energy-level differences are: $\Delta_{21} = 0.9581$, $\Delta_{31} = 1.8743$, $\Delta_{41} = 2.7486$, $\Delta_{32} = 0.9162$, $\Delta_{42} = 1.7905$, and $\Delta_{43} = 0.8743$, separately. They are mutually distinct so that Assumption 1 is satisfied.

The control goal is to reverse the population of initial state $\rho_0 = diag(0.3850, 0.2758, 0.1976, 0.1416)$, i.e., to drive the system to target state $\rho_f = diag(0.1416, 0.1976, 0.2758, 0.3850)$.

Firstly, consider a ladder-type system, in which the transitions between energy levels 1 and 2, 2 and 3, and 3 and 4, are permissible, as shown by the solid line in Fig. 1. There are three control Hamiltonians for the system:

$$H_1 = \begin{pmatrix} 0 & 1 & 0 & 0 \\ 1 & 0 & 0 & 0 \\ 0 & 0 & 0 & 0 \\ 0 & 0 & 0 & 0 \end{pmatrix}, H_2 = \begin{pmatrix} 0 & 0 & 0 & 0 \\ 0 & 0 & 1 & 0 \\ 0 & 1 & 0 & 0 \\ 0 & 0 & 0 & 0 \end{pmatrix}, H_3 = \begin{pmatrix} 0 & 0 & 0 & 0 \\ 0 & 0 & 0 & 0 \\ 0 & 0 & 0 & 1 \\ 0 & 0 & 1 & 0 \end{pmatrix} \quad (30)$$

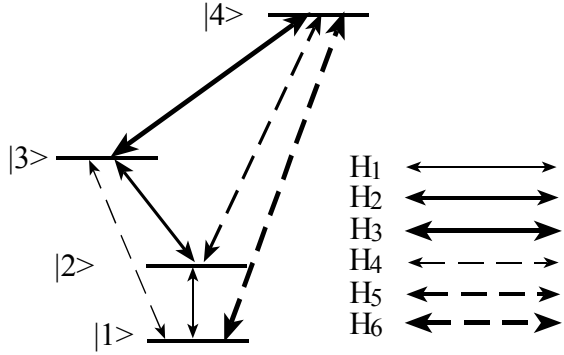

Fig. 1 The graph of the transitions among energy levels and the related control Hamiltonians

The control law is designed based on (5) and (30), in which the observable operator is constructed according to Theorem 1 and taken as $P = -\rho_f$ in simulation experiments; $\kappa_1 = \kappa_2 = \kappa_3 = 20$. In fact, it can be verified that the condition in theorem 2 is not satisfied in the case that control Hamiltonians are in the form (30). According to theorem 2, the $H_4$, $H_5$ and $H_6$ are added, and if the transitions between energy levels 1 and 3, 2 and 4, and 1 and 4 are permitted, as shown by the dashed line in Fig. 1, then three more control Hamiltonians are available.

$$H_4 = \begin{pmatrix} 0 & 0 & 1 & 0 \\ 0 & 0 & 0 & 0 \\ 1 & 0 & 0 & 0 \\ 0 & 0 & 0 & 0 \end{pmatrix}, H_5 = \begin{pmatrix} 0 & 0 & 0 & 0 \\ 0 & 0 & 0 & 1 \\ 0 & 0 & 0 & 0 \\ 0 & 1 & 0 & 0 \end{pmatrix}, H_6 = \begin{pmatrix} 0 & 0 & 0 & 1 \\ 0 & 0 & 0 & 0 \\ 0 & 0 & 0 & 0 \\ 1 & 0 & 0 & 0 \end{pmatrix} \quad (31)$$

Now, redesign control laws and do simulation experiments again with $\kappa_4 = \kappa_5 = \kappa_6 = 20$. Then, the evolution curves of the system population and the control fields are shown in Fig. 3, from which one can see that the controlled system reaches the target state $\rho_f$ after 100 a.u..

System simulation experiment results under the control strategy proposed is shown in Fig. 2 from which one cansee that although the extra control Hamiltonians are added, actually only two control fields $u_2$ and $u_6$ take effect. The control field $u_2$ reverses the population on energy levels 2 and 3, while $u_6$ reverses that on 1 and 4. These two control fields are similar to two $\pi$-pulses [13], by which the controlled system achieves the control goal. It also can be seen form Fig. 2 that control field $u_6$ vanished after 10 a.u., while $u_2$ holds longer. They are independent because they relate to different energy levels. Of course, one can reduce the effective time of $u_2$ by increasing $\kappa_2$.

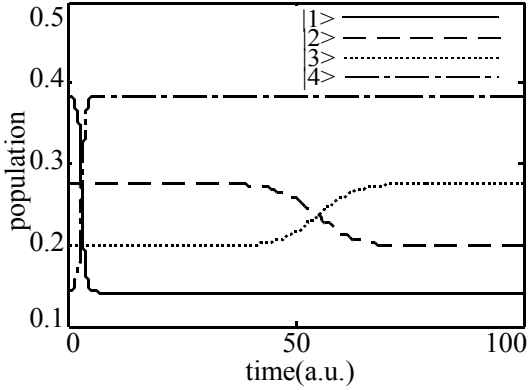  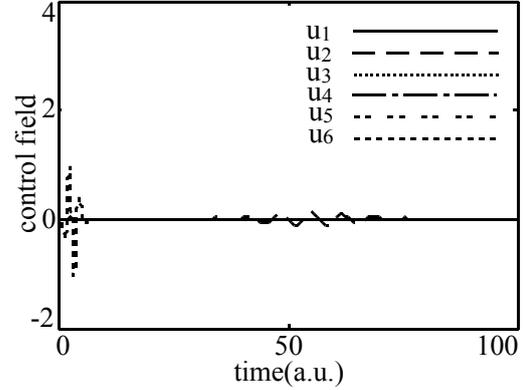

Fig. 2(a) Population of system          Fig. 2 (b) Control fields

Fig. 2 System simulation experiment results

## 6. Conclusions

In this paper, a convergent control strategy has been proposed for closed quantum systems in the interaction picture. We have expanded the Lyapunov function to an energy function by defining the observable operator with either positive or negative. The observable operator in the energy function has been derived such that any target state is the minimum point of the energy function, that is, the coherent vector of the observable operator must have opposite direction with that of the target state. The LaSalle invariant set has been given by introducing the Barbalat lemma, and dynamical stability of the states in the set has been analyzed, based on which, a sufficient and necessary condition on control Hamiltonians such that the system is convergent has been proven. At last, numerical experiments have been simulated on a four-level quantum system and demonstrated the effective results.

**Acknowledgements**

This work was supported partly by the National Key Basic Research Program under Grant No. 2011CBA00200.